\documentclass[12pt]{iopart}
\usepackage{epsfig}
\usepackage{amssymb}
\usepackage{color}
\usepackage{soul}

\usepackage{graphicx}
\begin{document}
\title[A geometric measure of non-classicality]
{A geometric measure of non-classicality}
\author{Paulina Marian and Tudor A. Marian}
\address{Centre for Advanced  Quantum Physics,
Department of Physics, University of Bucharest, 
R-077125 Bucharest-M\u{a}gurele, Romania}
\ead{paulina.marian@g.unibuc.ro}
\ead{tudor.marian@g.unibuc.ro}

\begin{abstract}
This paper aims to stress the role of the Cahill-Glauber quasi-probability densities in defining, 
detecting, and quantifying the non-classicality of field states in quantum optics.  The distance between 
a given pure state and the set of all pure classical states is called here a geometric degree 
of non-classicality.  As such, 
we investigate non-classicality  of a pure single-mode state of the radiation field by using 
the coherent states as a reference set of pure classical states. It turns out that any such distance is expressed in terms of the maximal 
value of the Husimi $Q$ function. As an insightful application we consider the de-Gaussification 
process produced when preparing a quantum state by adding $p$ photons to a pure Gaussian
one. For a coherent-state input, we get an analytic degree of non-classicality which compares 
interestingly with the previously evaluated entanglement potential. Then we show that addition 
of a single photon to a squeezed vacuum state causes a considerable enhancement 
of non-classicality, especially at weak and moderate squeezing of the original state. 
By contrast, addition of further photons is less effective.

\end{abstract}
\maketitle
\section{Introduction}
As first  pointed out by Glauber \cite{Gl}, there are states of the quantum radiation field  
for which  all the normally-ordered quantities are described by classical distributions. 
These states, which are now termed {\it classical}, have been singled out by Titulaer and 
Glauber \cite{TG} as possessing a well-behaved $P$ representation 
of the density operator  \cite{G2,S}, which is either a non-negative regular function or 
a distribution no more singular than Dirac's $\delta$.  The opposite situation, namely, 
the non-existence of the Glauber-Sudarshan P representation as a genuine probability density, 
is a largely accepted definition of non-classicality which generated a large amount of research 
in quantum optics  as can be seen in the survey Ref.\cite{D2002}.
  
The main concept we exploit in the present paper dedicated to an evaluation of non-classicality 
for pure states emerges from quantum information.  Accordingly, the distance from a given state 
having a specific property to a reference set of states not having it has been accepted  
as a measure of that property. It is notable that non-classicality of continuous-variable states 
was the first property proposed  to be quantified by a distance-type degree \cite{Hill1}:
\begin{eqnarray}
{\cal D}(\hat \rho):=\min_{{\hat \rho}^{\prime} \in {\cal C}}
d(\hat \rho, \hat {\rho}^{\prime}) 
\label{q},
\end{eqnarray}
where ${\cal C}$ is the convex set of all the classical states and $d$ is the distance between the density 
operators $\hat \rho$ and $\hat \rho^{\prime}$. However, the trace metric employed by Hillery 
in Ref. \cite{Hill1} and termed as {\em non-classical distance} turned out to be difficult to deal with analytically. Moreover, there is no parametrization of the whole set ${\cal C}$ of classical states to allow the extremization procedure required by Hillery's definition\ (\ref{q}). Subsequently, by restricting the set of all classical states 
to a tractable subset identified by a classicality criterion, and  by using more convenient metrics, 
the non-classical distance was successfully applied  to one-mode Gaussian states. Specifically, we mention 
the Hilbert-Schmidt metric used in \cite{DMW1,DMW2}, the Bures metric in \cite{PTH02}, 
the relative-entropy measure in Refs.\cite{PTH02,PTH04}  as well as the quantum Chernoff bound 
in Refs.\cite{Boca2009,Ghiu2010}. In our Refs.\cite{PTH02,PTH04}, the Bures degree of non-classicality 
for one-mode Gaussian states was found to fully agree with the earlier result of Lee's non-classical 
depth \cite{CTL}. 

Interest in non-classicality aspects has  recently renewed being stimulated by the ongoing resource
theories of various quantum properties \cite{Gour}.  Attempts to produce a resource theory 
of non-classicality \cite{GSV,Tan1,YBT} are in fact based on the identification of the set of classical 
states and classical operations. We point out that in Refs.\cite{PTH02, PTH04} co-authored 
with H.Scutaru, the present authors formulated a set of three requirements to make the definition\ (\ref{q})  acceptable as a measure of  non-classicality: 

C1) The degree of non-classicality vanishes if and only if the state 
is classical;

C2) Classical transformations preserve the degree of non-classicality;

C3) Non-classicality does not increase under any positive operator-valued measure (POVM).

We have there considered as being  {\it classical} those unitary transformations 
in Hilbert space which map coherent states into coherent states.  As such, the only one-mode 
classical transformations are the translations described by the displacement operators 
$D(\lambda)=\exp{(\lambda \hat a^{\dag}-\lambda^* \hat a)}$ and the rotations 
$R(\theta)={\rm exp}(-i\theta{\hat a}^{\dag}\hat a)$, written in terms of the amplitude operators 
of the mode $\hat a$ and $\hat a^{\dag}$. Therefore, the requirements C1)-C3) are quite similar 
to what is considered now to be necessary for defining a resource theory of non-classicality. 

In the present paper we restrict ourselves to investigate non-classicality of  pure continuous-variable 
states. What is really encouraging in defining a distance-type measure for pure states was discovered 
in the early days of quantum optics: a clear identification of the set of classical pure states.
Indeed, Cahill \cite{Cah} and later on Hillery \cite{Hill} proved that the only pure states that are classical 
are the coherent ones: all other pure (Gaussian and non-Gaussian) states are non-classical. Therefore, 
a geometric measure of non-classicality for a pure state $|\psi\rangle$ could be its distance to the set 
of all classical pure states, namely, the coherent ones. It appears to us that this measure  was first used 
in Ref.\cite{W} to quantify non-classicality of some popular pure states such as Fock states, 
squeezed vacuum, as well as even and odd coherent states. Soon after, in  Ref. \cite{MB}  
a comparison between this distance-type measure and Lee's non-classical depth for pure states 
has revealed an advantage of the geometric degree which turned out to be more sensitive. 
 
The plan of our paper is as follows. In section 2 we discuss the role  played by  the quasi-probability 
distributions introduced by Cahill and Glauber \cite{CG}  in describing and measuring non-classicality.
Section 3 investigates first the non-classicality of an interesting pure state  which is
important for experiments: a coherent  state with $p$ added-photons \cite{AT1}.  We give here 
an analytic form of the defined geometric degree of non-classicality and compare it with some previously 
used indicators of non-classicality.  Another interesting example we are interested in is provided 
by the squeezed vacuum states (SVSs) whose non-classicality enhancement by addition of photons
is here analysed. The concluding section 4 stresses the usefulness of the geometric measure 
and its consistency with other evaluations of the amount of non-classicality.

\section{Non-classicality and quasi-probability distributions}
In continuous-variable settings,  phase-space formalism allows us to conveniently describe 
quantum states using the generalised quasi-probability distributions
\cite{CG}:
\begin{eqnarray}
W(\beta,s)=\frac{1}{\pi}\int {\rm d}^2\lambda \exp{(\beta\lambda^*-\beta^*\lambda)} \;\chi(\lambda,s),
\label{ws}
\end{eqnarray}
which are in fact the Fourier transforms of the $s$-ordered characteristic functions (CFs) 
of the density operator $\hat \rho$, 
\begin{eqnarray}
\chi(\lambda,s):=\exp{(\frac{s}{2}|\lambda|^2)}{\rm Tr} [\hat \rho D(\lambda)].
\label{cf}
\end{eqnarray}
The most significant quasi-probability distributions arise from the normally ordered CF 
($ s=1, \,\frac{1}{\pi} W(\beta,1) =:P(\beta)$ is the Glauber-Sudarshan $P$ function), 
symmetrically ordered CF($s=0, \, W(\beta,0)=:W(\beta)$ is the  Wigner function), 
and anti-normally ordered CF  ($ s=-1, \,\frac{1}{\pi} W(\beta,-1) =:Q (\beta)$ is the Husimi 
$Q$ function).

It is quite interesting that all the three above-defined quasi-probability distributions play a role 
in describing non-classicality, especially in the pure-state case. First, the behaviour 
of the $P$ representation {\em defines} the concept of non-classicality in quantum optics: 
non-classical states are those having a $P$ representation which is negative or more singular 
than Dirac's $\delta$. Second,  according to the Hudson's theorem \cite{H}, only the pure Gaussian 
states possess a non-negative Wigner function, all the other non-Gaussian pure states display 
negativities of their Wigner distribution.  Signatures of non-classicality could  be thus identified
through the negativity of the Wigner function, for pure and even some mixed non-Gaussian 
states \cite{ZK}. Therefore, the Wigner function {\em detects} non-classicality of all pure states 
except for the Gaussian ones. Finally, to see the role played by the $Q$ function in describing
non-classicality, let us define the geometric degree of non-classicality of a pure one-mode state 
$|\Psi\rangle$:
\begin{eqnarray}
{\cal D}_{Q}:=1-\max_{\{\beta\}}|\langle \beta|\Psi\rangle|^2, 
\label{1}
\end{eqnarray}
where ${\{|\beta\rangle\}}$ is the set of the coherent states. The following remarks are at hand 
when looking at the definition\ (\ref{1}):
\begin{enumerate}
\item ${\cal D}_{Q}$ is an exact distance-type measure for pure states because the reference set 
of classical pure states is exhaustive.
\item ${\cal D}_{Q}$ is defined in terms of the maximum of the $Q$ function,
\begin{eqnarray}
Q(\beta):=\frac{1}{\pi}\langle \beta |\Psi \rangle \langle \Psi |\beta\rangle.
\label{Husimi}
\end{eqnarray}

\item ${\cal D}_{Q}$ obviously meets the requirements  C1)-C2): 

 $0\leqq  {\cal D}_{Q}\leqq 1$, $\; {\cal D}_{Q}=0\;\;{\rm for \; coherent \;states}$;
 
 ${\cal D}_{Q}$ is invariant under displacements and rotations.
 
\end{enumerate}
To conclude, the maximum of the $Q$ function is a reliable {\em measure} of non-classicality.
The advantages of a geometric degree of non-classicality for pure states were recognized long ago. 
${\cal D}_{Q}$ was first evaluated in some simple cases  including Gaussian pure states \cite{W}.
For non-Gaussian pure states, some examples were examined \cite{MB} showing that the distance-type 
measures are more appropriate to quantify non-classicality than the non-classical depth. 
Recently, in Ref.\cite{Nair},  improved upper and lower bounds on Hillery's non-classical 
distance were written in terms of the $Q$ function of the given state.
 
\section{Exactly solvable examples: photon-added pure Gaussian states}

Let us consider the output state generated by the addition of $p$ photons to an arbitrary 
pure state $|\Psi_0\rangle$ : 
\begin{eqnarray}
|\Psi_p \rangle:={\cal N}_p (\hat a^{\dag})^p | \Psi_0 \rangle,        \qquad  (p=1, 2, 3, ...),
\label{PAS}
\end{eqnarray}
where ${\cal N}_p$ is a normalization factor whose squared modulus reads: 
\begin{eqnarray}
 |{\cal N}_p|^2=\left[ \langle  \Psi_0|({\hat a}^p(\hat a^{\dag})^p| \Psi_0 \rangle \right]^{-1}.
\label{N}
\end{eqnarray}
Accordingly, the normalization constant is determined by the $p$th-order antinormally-ordered 
correlation function of the input state $| \Psi_0 \rangle$. Further, by employing in Eq.\ (\ref{Husimi}) 
the eigenvalue equation of the annihilation operator, $\hat a |\beta \rangle= \beta |\beta \rangle$,  
we find that the $Q$ function of the output state\ (\ref{PAS}) is proportional to that of the input state 
$| \Psi_0 \rangle$:
\begin{eqnarray} 
Q_p(\beta)=|{\cal N}_p |^2  |\beta |^{2 p} Q_0(\beta),          \qquad  (p=1, 2, 3, ...).
\label{Q}
\end{eqnarray}

As we know that the $Q$ function for a Gaussian state \cite{PT93} is an exponential of the variables 
$\beta$ and $\beta^*$, Eq.\ (\ref{Q}) shows us that addition of photons is a non-Gaussian operation. 
Moreover, according to a theorem of Lee, by removing the vacuum from the Fock-state 
expansion of a classical state, we get a non-classical output \cite{Lee95}. 
This can be done by adding photons to a classical state such as a coherent one \cite{AT1} 
or a thermal one \cite{AT2}. We can say that photon-added states are both non-classical 
and non-Gaussian. This is important because the non-Gaussianity of states and operations turned out 
to be an indispensable  property for some protocols in quantum information with continuous variables 
such as: entanglement distillation \cite{PE2002}, as well as distillation of any other 
quantum resource \cite{Winter}, error correction \cite{NFC}, and universal quantum 
computation \cite{BL,QC}. For the sake of simplicity and in view of physical relevance, we chose 
to discuss here non-classicality for two classes of states obtained by addition of photons 
to pure Gaussian ones:  photon-added coherent states and photon-added SVSs.
 
The $p$-photon-added coherent state, 
\begin{eqnarray}
|\Psi_p(\alpha) \rangle:= \frac{1}{[p!L_p(-|\alpha|^2)]^{1/2}}(\hat a^{\dag})^p | \alpha \rangle,  
\label{PAC}
\end{eqnarray}
was formally defined  and characterized from the quantum optical perspective in Ref.\cite{AT1}. 
It was found that its $P$ distribution is highly singular and its Wigner function displays negative values. 
Other non-classical properties such as sub-Poissonian statistics and amplitude squeezing  were also 
analysed. Note that in Eq.\ (\ref{PAC}),  $L_p(u)$ is a Laguerre polynomial, which is always positive 
for negative values of its argument, as shown by Eqs. 10.12 (7) and/or 10.12(14) in Ref.\cite{HTF2}.

A significant piece of progress was the experimental preparation and investigation 
of a single-photon-added coherent state by Zavatta {\em et al.} \cite{Z1,Z2}. Remark that 
single-photon-added coherent states are interesting from a fundamental point of view 
as they represent the result of the simplest excitation of a classical light field. It can be generated 
by injecting a coherent state $|\alpha \rangle$ into the signal mode  of an optical parametric amplifier 
and a conditioning of the state based on measurements on the idler mode to select the one-photon 
excitation. Moreover, the experiment allowed the tomographic reconstruction of the Wigner function. 
Its negativity was detected and analysed while  a significant degree of squeezing was also found. 

Let us now proceed to find ${\cal D}_{Q}$ defined in Eq.\ (\ref{1}) for the state\ (\ref{PAC}). 
The {$Q$ function for $|\Psi_p\rangle$} is \cite{AT1}:
\begin{eqnarray}  
Q_p( \alpha, \beta)=\frac{1}{\pi} |\langle \beta|\Psi_p (\alpha)\rangle|^2=
\frac{1}{\pi}\frac{|\beta|^{2 p}}{p!L_p(-|\alpha|^2)}\exp{(-|\beta-\alpha|^2)},
\label{QF} 
\end{eqnarray}
The single-peaked aspect of  $Q_p$ can be seen in Fig.1 for different numbers of added photons 
to the same coherent state.

\begin{figure*}[t]
\center
\includegraphics[width=7cm]{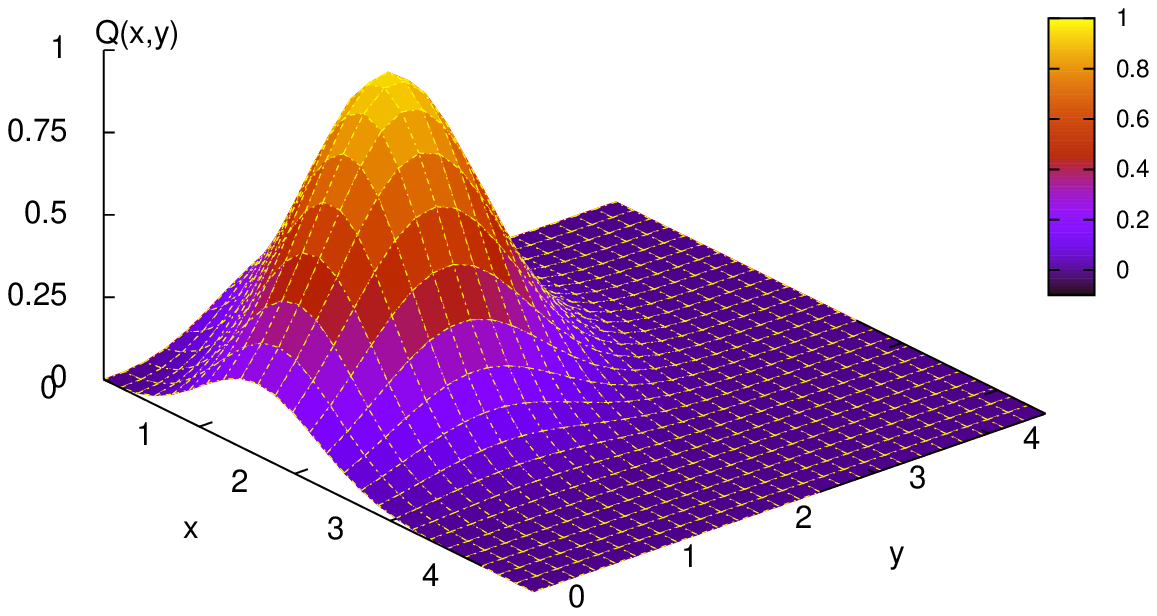}
\includegraphics[width=7cm]{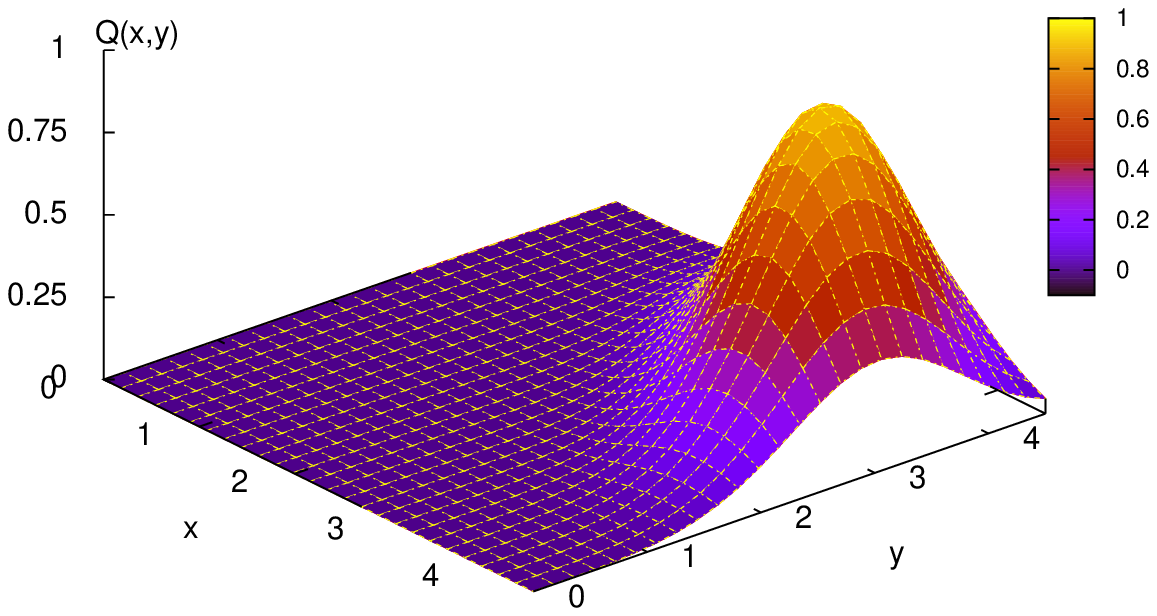}
\caption{The $Q$ function of a $p$-photon-added coherent state. The parameters of the state are: 
$p=1, \Re e (\alpha)=\Im m (\alpha)=2$ (top) and  $p=10, \Re e (\alpha)=\Im m (\alpha)=2$ (bottom). 
For the sake of simplicity, we have denoted $x:=\Re e (\beta), y:=\Im m (\beta)$.}
\end{figure*}
Maximization of $Q_p(\beta)$ with respect to the variables $\Re e (\beta)$ and $\Im m (\beta)$ 
is routinely performed to nicely give a general formula:
\begin{eqnarray}
Q_p^{max}(\alpha)&=&\frac{1}{\pi}\frac{1}{p!L_p(-|\alpha|^2)}\left[\frac{|\alpha|}{2}(\sqrt{1+4 p/|\alpha|^2}+1)
\right]^{2p}\nonumber\\ &&\times\exp{\left[-\frac{|\alpha|^2}{4}(\sqrt{1+4 p/|\alpha|^2}-1)^2\right]}.
\label{qm}
\end{eqnarray}
The geometric degree of non-classicality  $ {\cal D}_Q(|\Psi_p(\alpha) \rangle \langle \Psi_p(\alpha)|)
:=1-\pi Q_p^{max} (\alpha)$ depends therefore only on the number $p$ of added photons 
and the mean occupancy of the coherent state $\langle \hat a^{\dag} \hat a \rangle=|\alpha|^2$. 
Plots of ${\cal D}_Q$ with respect to the number $p$ of added photons and  the coherent mean 
occupancy $|\alpha|^2$  in Fig.2 show that non-classicality decreases drastically with the intensity 
of the coherent beam and increases with the number of added photons.
The incipient values on the plots of Fig.2b are the degrees of non-classicality of the Fock 
states with $p=1,5,10$, respectively. When taking the limit $\alpha =0$ in Eq.\ (\ref{qm}) 
we get the maximal $Q$ function for the number state $|p \rangle$:
\begin{equation}
 Q_p^{max}(0)=\frac{1}{\pi}\frac{1}{p!}\left( \frac{p}{\rm e} \right)^p,   \qquad  (p=1, 2, 3, ...).
\label{Qp}
\end{equation}
Equation (\ref{Qp}) coincides with a directly obtained formula for the maximal $Q$ function of a Fock state.
Let us write the corresponding degree of non-classicality of a Fock state:
\begin{equation}
{\cal D}_Q(|p \rangle \langle p |)=1-\frac{1}{p!}\left( \frac{p}{\rm e} \right)^p,   \qquad  (p=1, 2, 3, ...).
\label{DQp}
\end{equation} 
In view of the inequality
\begin{equation}
\frac{Q_{p+1}^{max}(0)}{Q_p^{max}(0)}=\frac{1}{\rm e}\left( 1+\frac{1}{p} \right)^p <1,
 \qquad  (p=1, 2, 3, ...),
\label{u_p}
\end{equation} 
the sequence\ (\ref{Qp}) is strictly decreasing and therefore the geometric degree 
of non-classicality\ (\ref{DQp}) increases with the number of  photons. For large 
photon numbers,  we write Eq.\ (\ref{DQp}) using Stirling's approximation,  
Eq. 1.18 (3) in Ref.\cite{HTF1}:
\begin{equation}
{\cal D}_Q(|p \rangle \langle p |) \thickapprox 1-\frac{1}{\sqrt{2\pi p} },   \qquad  (p \gg 1).
\label{DQP}
\end{equation} 

\begin{figure*}[h]
\center
\includegraphics[width=7cm]{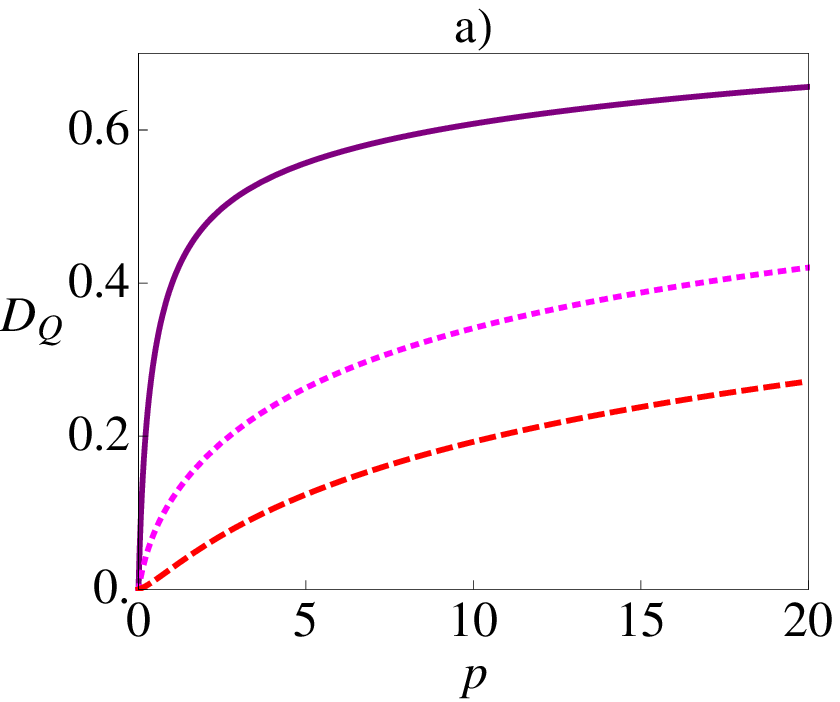}
\includegraphics[width=7cm]{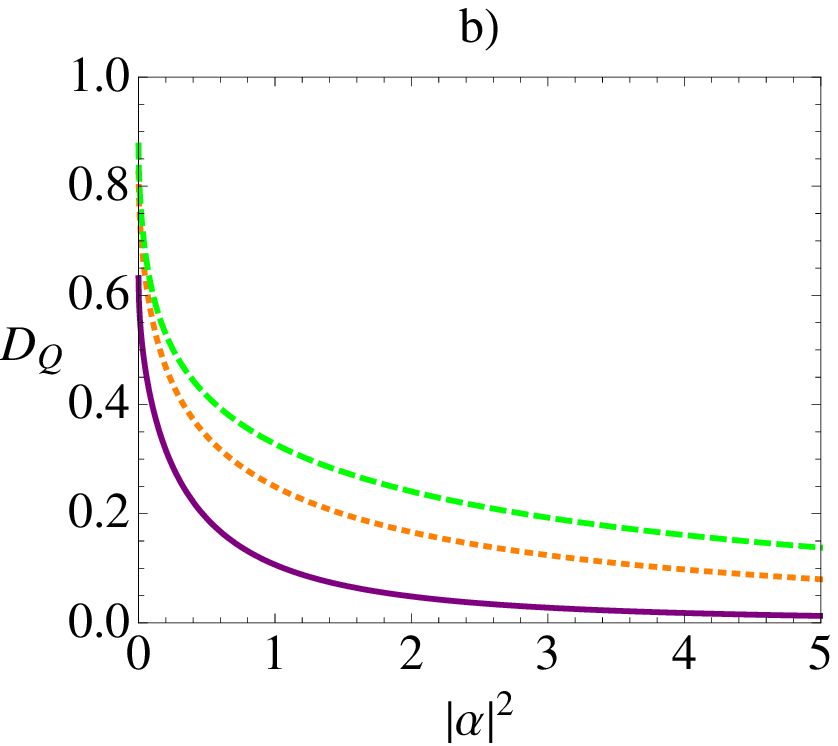}
\caption{Photon-added coherent states: a) ${\cal D}_Q$ increases with the number $p$ of added photons.
The coherent mean occupancies are $|\alpha|^2=3$ (red dashed plot); $0.9$ (magenta dotted curve) 
and  $0.1 $ (purple line).   b) Non-classicality degree is  much larger for smaller coherent mean occupancies. 
We have used  $p=1$ (purple plot), $p=5$ (red dotted plot) and $p=10$ (green dashed curve).}
\end{figure*}
According to Fig. 2,  non-classicality degrees ${\cal D}_Q$  converge to $0$ for higher coherent
intensities $|\alpha|^2$, but are well separated for low beam intensity,
being larger  for higher numbers of photons added  to a weak coherent state. 

Non-classicality of a $p$-photon-added coherent state was discussed in Ref.\cite{Usha}
by evaluating its entanglement potential (EP). This measure of non-classicality is defined 
as the entanglement generated by a non-classical state  when mixed with the vacuum state 
in a balanced beam splitter \cite{Asboth}. As such, the task of measuring non-classicality is transferred 
to an evaluation of a two-mode entanglement which is not a simpler problem in most cases of interest.
Fortunately, for some lower numbers of added photons to a coherent state the EP could be evaluated 
in Ref.\cite{Usha}. This gives us the opportunity to compare our present distance-type results 
to those given by a non-classicality measure of a totally different origin, the EP.  Our formula\ (\ref{qm}) 
depicted in Fig. 2  describes the behaviour of ${\cal D}_Q$ in close agreement with the investigation 
in Ref.\cite{Usha}. It is sufficient to compare our Fig.2 b) and the corresponding Fig.3  in Ref.\cite{Usha} 
to see their similarity. We have to remark that only for single-photon-added coherent states was possible 
to have an analytic EP in \cite{Usha}. Even for $p=2$ the EP-evaluation was performed only numerically.
Other findings in Ref.\cite{Usha} based on observing the negativity of the Wigner function enhance 
the idea of consistency between ${\cal D}_Q$ and other indicators of non-classicality. 

The second example we want to address now is the modification of non-classicality by adding photons 
to a non-classical state, namely an SVS, $| \Psi_{SV}(r,  \varphi)  \rangle=| \Psi_0(r,  \varphi) \rangle$;
its $Q$ function has the explicit expression \cite{PT93}:
\begin{eqnarray} 
Q_0(r, \varphi, \beta)=\frac {1}{\pi \cosh r}\exp{\{-|\beta|^2
\left[ 1-(\tanh r) \cos(\varphi-2 {\rm arg}\;(\beta)) \right]\}},
\label{q0}
\end{eqnarray}
where $r$ is the squeeze parameter and $\varphi$ is the squeeze angle. 
Notice first that the maximum of the $Q$ function for an SVS  is reached for 
$\beta_{max}= 0$ and has the value  
\begin{equation}
Q_0^{max}(r)=(\pi \cosh r)^{-1}. 
\label{Q_0}
\end{equation} 
Further, the maximum 
of $Q_p (\beta)$, Eq.\ (\ref{Q}),  is found to be at $ | \beta_{max} |^2=p\;  {\rm e}^r \cosh r $ 
and ${\rm arg} ({\beta_{max}})=\frac{1}{2}\varphi$. 
The next step to apply Eq.\ (\ref{Q}) is to evaluate the $p$th-order antinormally-ordered correlation 
function for an SVS. Its derivation parallels that of the normally-ordered one in Ref.\cite{P91}. 
We find the general formula:
\begin{eqnarray} 
\langle  \Psi_{SV}(r,  \varphi) |{\hat a}^p(\hat a^{\dag})^p | \Psi_{SV}(r,  \varphi) \rangle
=p! \, (\cosh r)^{2 p}\, {_{2}F_{1}} \left( -\frac{p}{2}, -\frac{p-1}{2}; 1; (\tanh r )^2 \right), 
\nonumber\\     (p=0, 1, 2, 3, ...).
\label{anti} 
\end{eqnarray}
The above Gauss hypergeometric function ${_{2}F_{1}}$  is a polynomial in the variable 
$\tanh r$, of degree either $p$ if $p$ is even or $p-1$ if $p$ is odd, as seen from Eq. 2.1(2) 
in Ref. \cite{HTF1}.  We finally get the ratio of maximal values 
\begin{eqnarray} 
\frac{Q_p^{max}(r)}{Q_0^{max}(r)} = \frac{1}{p!}\left(\frac{ p\, e^{r-1}}{ \cosh r}\right)^p
\left[{_{2}F_{1}} \left( -\frac{p}{2}, -\frac{p-1}{2}; 1; (\tanh r )^2 \right)\right]^{-1}.
\label{qsv} 
\end{eqnarray}
By specializing Eq.\  (\ref{qsv}) to the simplest cases $p=1$ and $p=2$, we find the inequalities:
\begin{eqnarray} 
Q_1^{max}(r)=\frac{2} {\rm e}\, \frac{1}{1+\exp(-2r)}\, Q_0^{max}(r) < Q_0^{max}(r);
\label{p=1}
\end{eqnarray}
\begin{eqnarray} 
Q_2^{max}(r)=\frac{16}{3}\frac{1}{ {\rm e}^2}\, \frac{1}{1+\frac{2}{3}\exp(-2r)+\exp(-4r)}
\, Q_0^{max}(r) <Q_0^{max}(r).
\label{p=2}
\end{eqnarray}
Both Eqs.\ (\ref{p=1}) and\ (\ref{p=2}) show that non-classicality is enhanced by adding photons 
to this non-classical state. 

Nevertheless, it is instructive to analyse the general formula\ (\ref{qsv}) in two limit cases: 
zero squeezing $(r=0)$ and very strong squeezing $(r \to \infty)$. The first one reduces 
to that of the Fock states, already discussed above [Eqs.\ (\ref{Qp})-\ (\ref{DQP})]. In order to get 
a convenient expression valid in the strong-squeezing regime, we employ Eq.\ (\ref{qsv})  
in conjunction with two well-known identities,  Eqs. 2.1(14) and 1.2(15) in Ref. \cite{HTF1}:  
Gauss's summation formula,
\begin{eqnarray} 
_{2}F_{1}(a, b; c; 1)=\frac{\Gamma(c)\, \Gamma(c-a-b)}{\Gamma(c-a)\, \Gamma(c-b)}\, ,   
\nonumber\\         (c \ne 0, -1, -2, -3, ..., \;  \Re(c-a-b)>0),
\label{Gauss}
\end{eqnarray}
and, respectively,  Legendre's duplication formula,
\begin{eqnarray} 
2^{2z-1}\Gamma(z)\, \Gamma\left( z+\frac{1}{2} \right)={\pi}^{\frac{1}{2} }\, \Gamma(2z).
\label{Legendre}
\end{eqnarray}
We find the limit ratio
\begin{eqnarray} 
\lim_{r \to \infty} \frac{Q_p^{max}(r)}{Q_0^{max}(r)}=\left( \frac{p}{\rm e} \right)^p
\frac{\sqrt{\pi} }{\Gamma\left( p+\frac{1}{2} \right)}\, ,  \qquad  (p=1, 2, 3, ...),
\label{p/0}
\end{eqnarray}
as well as a stronger inequality than Eq.\ (\ref{u_p}):
\begin{eqnarray} 
\lim_{r \to \infty} \frac{Q_{p+1}^{max}(r)}{Q_p^{max}(r)}=\frac{1}{\rm e}\left( 1+\frac{1}{p} \right)^p
\frac{p+1}{p+\frac{1}{2} }<1,    \qquad    (p=1, 2, 3, ...).
\label{v_p}
\end{eqnarray} 
Accordingly, for $r \to \infty$, the sequence $\{ Q_p^{max}(r) \}$ strictly decreases to a vanishing limit.
For large photon numbers, we apply again Stirling's approximation, 
Eq. 1.18 (3) in Ref. \cite{HTF1}, to get an asymptotic expression of Eq.\ (\ref{p/0}): 
\begin{eqnarray} 
\lim_{r \to \infty} \frac{Q_p^{max}(r)}{Q_0^{max}(r)}=\frac{1}{\sqrt{2} }\, ,  \qquad  (p \gg 1).
\label{pgg1}
\end{eqnarray} 
In conclusion, in the strong-squeezing regime $(r \to \infty)$, the geometric degree of non-classicality 
${\cal D}_Q(|\Psi_p(r,  \varphi) \rangle \langle \Psi_p(r,  \varphi)|)$ strictly increases with the number 
of added photons reaching a limit equal to unity.

\begin{figure*}[h]
\center
\includegraphics[width=7cm]{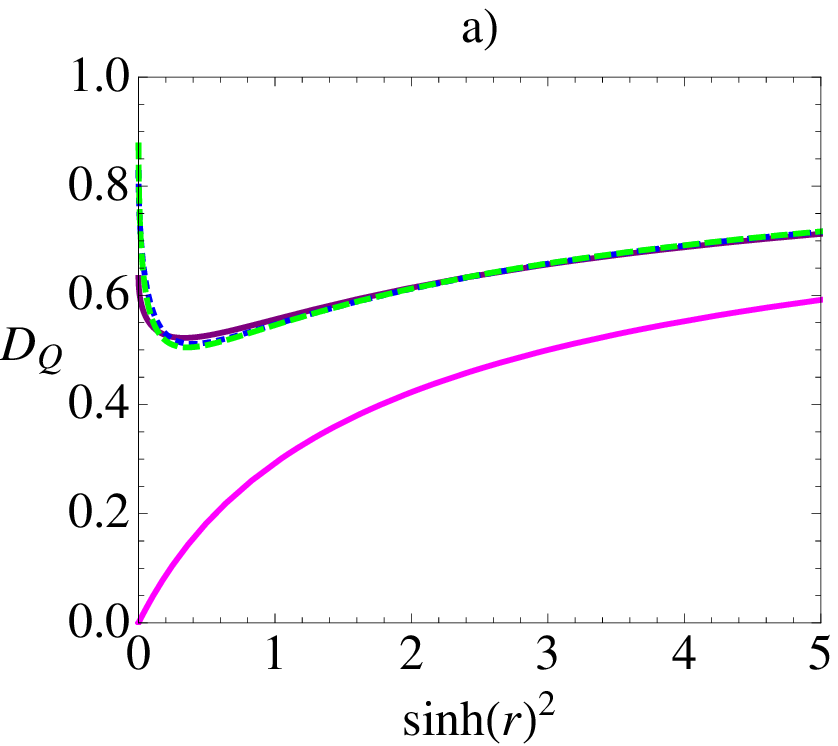}
\includegraphics[width=7cm]{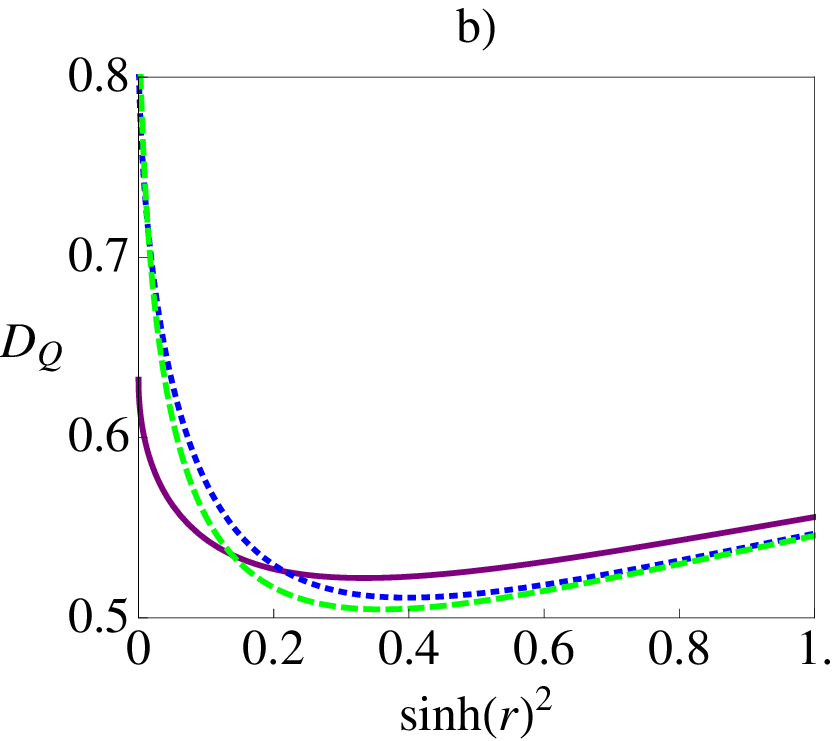}
\caption{Photon-added squeezed vacuum states: a) ${\cal D}_Q$ has a minimum at small  mean 
occupancies $\langle \hat a^{\dag} \hat a\rangle=(\sinh r)^2$ of the input SVS. At larger squeezing input, 
its increase is monotonic and very weakly dependent on the number $p$ of added photons.
The magenta plot starting from origin is the non-classicality degree ${\cal D}_Q$ of the input SVS
which is much lower than all the corresponding added-photon ones. However, the degrees of
non-classicality ${\cal D}_Q$ of the original SVS and its $p$-photon added relatives tend 
to the maximal value $1$ 
for strong squeezing. b)  A better view of the minimum at weak input squeezing. We have used 
$p=1$ (purple plot), $p=5$ (blue dotted plot), and $p=10$ (green dashed curve).}
\end{figure*}
However, Figure 3 displays an interesting feature of the degree of non-classicality 
${\cal D}_Q(|\Psi_p(r,  \varphi) \rangle \langle \Psi_p(r,  \varphi)|)$ as a function of the mean photon 
number $\langle \hat a^{\dag} \hat a\rangle=(\sinh r)^2$ in the input SVS. The graph for any number 
of added photons has a minimum for a rather small value of $r$ which essentially separates
the ranges of weak and strong squeezing. Although in both limit cases of very weak and very strong 
squeezing the hierarchy in the number of added photons is strictly observed, this does not happen 
for intermediate squeezing, i. e., in the neighbourhood of the minima, where various graphs cross
each other. It is worth to point out a fact which is important from an experimental perspective. 
Addition of a single photon to an SVS provides a substantial enhancement of non-classicality 
at weak and moderate squeezing, while addition of a larger number of photons does not make 
any  significant difference in non-classicality.  

\section{Conclusions}

The use of the quasi-probability densities introduced by Cahill and Glauber \cite{CG} 
has a long and successful  history in quantum optics. The present paper emphasizes
their special significance for non-classicality matters.  Thus the P representation is essential 
in {\em defining} the non-classicality concept, the Wigner function {\em detects} non-classicality 
by its negativity,  and the Q function is  a proper {\em quantifier} for the non-classicality of pure states. 
When dealing with pure states, all the distance-type measures of non-classicality defined 
with respect to the unique set of pure classical states, namely,  the coherent ones, are expressed 
in terms of maximal value of the $Q$ function of the given state. This happens because, 
in the pure-state case, all the metrics having the suitable abilities to define a distance-type degree, 
such as Hilbert-Schmidt, Hellinger, and  Bures metrics, depend only on the $Q$ function of the state. 
We can say that ${\cal D}_Q$ is a properly defined measure of non-classicality 
because the reference set of classical coherent states is exhaustive. Interestingly, 
a similar geometric measure  was proposed long ago by Shimony for the  entanglement 
of pure two-mode states \cite{Shimony}. The chosen reference set was there composed 
of all pure product states. According to recent research on non-classicality \cite{Winter,YBT,Gour} 
and non-Gaussianity \cite{TZ}, some protocols in quantum information processing perform better 
when using non-Gaussian resources and operations. This is a good reason for a future extension 
of non-classicality quantification to a larger sector of non-Gaussian pure states such as 
photon-added generalized coherent states \cite{Tav,Moja,Hussin}.

As applications of quantifying non-classicality of pure states we have here considered  states 
obtained by adding photons to Gaussian states. Our Eqs.\ (\ref{qm}) for photon-added coherent states 
and\ (\ref{qsv}) for photon-added SVSs are quite remarkable because an analytic degree of non-classicality 
is difficult to obtain for non-Gaussian states, which have in general a more complicated structure. 
The states we have dealt with in this paper are important from the experimental feasibility as well.  
Along these lines we have found that a coherent state of weak intensity gains more non-classicality 
by addition of photons than a strong-intensity one. We have also shown that even when a single photon 
is added to an SVS, this considerably enhances the amount of non-classicality of the input state 
especially at weak values of squeezing.

\ack{
 This work was supported by the funding agency CNCS-UEFISCDI of the Romanian Ministry 
 of Research and Innovation through grant No. PN-III-P4-ID-PCE-2016-0794.}


\newpage
\section*{References}

\begin{thebibliography}{70}
\bibitem{Gl}  Glauber R J  1963 Coherent and Incoherent States of the Radiation Field  
{\em Phys. Rev.} {\bf 131}  2766 
\bibitem{TG}  Titulaer U M and Glauber R J 1965 Correlation Functions for Coherent Fields 
{\em Phys. Rev.} {\bf 140}  B676
\bibitem{G2} Glauber R J  1963  Photon Correlations {\em Phys. Rev. Lett.}  {\bf 10} 84 
\bibitem{S} Sudarshan E C G 1963 Equivalence of Semiclassical and Quantum Mechanical 
Descriptions of Statistical Light Beams {\em Phys. Rev. Lett.}  {\bf 10} 277 
\bibitem{D2002}Dodonov V V 2002 'Nonclassical'  states in quantum optics: a 'squeezed' review 
of the first 75 years {\em J. Opt. B: Quantum Semiclass. Opt.}  {\bf 4 } R1
\bibitem{Hill1}  Hillery M 1987 Nonclassical distance in quantum optics {\em  Phys. Rev. A}  {\bf 35} 725 
\bibitem{DMW1} Dodonov V V, Man'ko O V, Man'ko V I,   W\"unsche  A  1999 Energy-sensitive 
and 'classical-like' distances between quantum states {\em Phys. Scr.}  {\bf 59} 81
\bibitem{DMW2}Dodonov V V, Man'ko O V, Man'ko V I,   W\"unsche  A  2000 Hilbert-Schmidt distance 
and nonclassicality of states in quantum optics {\em J. Mod. Opt. } {\bf 47} 633
\bibitem{PTH02}Marian P, Marian T A and Scutaru H 2002 Quantifying Nonclassicality of One-Mode
Gaussian States of the Radiation Field  {\em Phys. Rev. Lett.} {\bf 88} 153601 
\bibitem{PTH04}Marian P, Marian T A and Scutaru H 2004  Distinguishability and nonclassicality 
of one-mode Gaussian states {\em Phys. Rev. A} {\bf 69}  022104 
\bibitem{Boca2009}  Boca M, Ghiu I, Marian P, Marian T A  2009 Quantum Chernoff bound 
as a measure of nonclassicality for one-mode Gaussianstates  {\em Phys. Rev. A }  {\bf 79}  014302 
\bibitem{Ghiu2010}Ghiu I, Bj\"ork G, Marian P, Marian T A  2010 Probing light polarization 
with the quantum Chernoff bound  {\em Phys. Rev. A}  {\bf 82} 023803 
\bibitem{CTL} Lee C T 1991 Measure of the nonclassicality of nonclassical states 
{\em  Phys. Rev. A } {\bf 44} R2775
\bibitem{Gour}  Chitambar  E and Gour G  2019  Quantum resource theories
{\em Rev. Mod. Phys.}  {\bf 91} 025001 
\bibitem{GSV}  Gehrke C, Sperling J, and Vogel W 2012 Quantification of nonclassicality 
{\em Phys. Rev. A } {\bf 86} 052118 
\bibitem{Tan1} Tan K C,  Volkoff T,  Kwon H, and Jeong H 2017 Quantifying the Coherence 
between Coherent States  {\em Phys. Rev. Lett.} {\bf 119} 190405 
\bibitem{YBT}  Yadin B,  Binder F C, Thompson J,  Narasimhachar V, Gu M, and 
Kim M S 2018  Operational Resource Theory of Continuous-Variable Nonclassicality 
{\em  Phys. Rev. X }{\bf 8} 041038 
\bibitem{Cah}Cahill K E 1969 Pure States and the P Representation {\em  Phys. Rev.}  {\bf 180} 1239 
\bibitem{Hill} Hillery M  1985 Conservation laws and nonclassical states in nonlinear optical systems, 
{\em Phys. Rev. A}  {\bf 31} 338 
\bibitem{W}W\"unsche  A , Dodonov V V, Man'ko O V, Man'ko V I  2001
Nonclassicality of states in quantum optics {\em  Fortschr. Phys.} {\bf 49}  1117 
\bibitem{MB}  Malbouisson J M C and  Baseia B 2003 On the Measure of Nonclassicality 
of Field States {\em Phys. Scr.}  {\bf 67} 93 
\bibitem{CG} Cahill K E and  Glauber R J  1969 Density Operators and Quasiprobability 
Distributions {\em Phys. Rev.} {\bf 177} 1882 
\bibitem{AT1} Agarwal G S and Tara K  1991 Nonclassical properties of states generated 
by the excitations on a coherent state {\em Phys. Rev. A}  {\bf 43}, 492 
\bibitem{H} Hudson R L 1974  When is the Wigner quasi-probability density non-negative? 
{\em  Rep. Math. Phys.} {\bf  6} 249
\bibitem{ZK} Kenfack A and \.{Z}yczkowski K 2004  Negativity of the Wigner function as an indicator 
of non-classicality  {\em J. Opt. B: Quantum Semiclass. Opt.}  {\bf 6 } 396
\bibitem{Nair} Nair R 2017 Nonclassical distance in multimode bosonic systems
 {\em Phys. Rev. A} {\bf  95} 063835
 \bibitem{PT93} Marian P and  Marian T A  1993 Squeezed states with thermal noise. 
I. Photon number statistics {\em Phys. Rev. A } {\bf 47} 4474
\bibitem{Lee95}  Lee C T  1995 Theorem on nonclassical states {\em Phys. Rev. A}  {\bf 52} 3374 
\bibitem{AT2}  Agarwal G S and Tara K  1992
Nonclassical character of states exhibiting no squeezing or sub-Poissonian statistics  
{\em Phys. Rev. A}  {\bf 46} 485
\bibitem{PE2002}Eisert J, Scheel S, and  Plenio M B 2002 Distilling Gaussian States with Gaussian 
Operations is Impossible  {\em Phys. Rev. Lett.} {\bf 89} 137903
\bibitem{Winter}Lami L,  Regula B,  Wang X,  Nichols R,  Winter A, and  Adesso G 2018 Gaussian 
quantum resource theories {\em Phys. Rev. A } {\bf 98} 022335 \bibitem{NFC}  Niset J,  Fiur\'a\u sek J, 
and Cerf N J 2009 No-Go Theorem for Gaussian Quantum Error Correction 
{\em Phys. Rev. Lett.} {\bf 102} 120501 
\bibitem{BL} Lloyd S and  Braunstein S L 1999 Quantum Computation over Continuous Variables
{\em Phys. Rev. Lett.} {\bf  82} 1784 
\bibitem{QC} Menicucci N C, van Loock P, Gu M,  Weedbrook C,  Ralph T C, and  Nielsen M A 2006 
Universal Quantum Computation with Continuous-Variable Cluster States
{\em Phys. Rev. Lett.} {\bf 97} 110501 
\bibitem{HTF2} Erd\'{e}lyi A,  Magnus W,  Oberhettinger F, and Tricomi F G  1953 
{\em Higher Transcendental Functions} (New York: McGraw--Hill), Vol. 2.
\bibitem{Z1} Zavatta A, Viciani S, Bellini M 2004
Quantum-to-classical transition with single-photon-added coherent states of light 
{\em Science}  {\bf 306} 660
\bibitem{Z2} Zavatta A, Viciani S, Bellini M 2005 Single-photon excitation of a coherent state:Catching the elementary step of stimulated light emission {\em Phys. Rev. A } {\bf 72} 023820
\bibitem{HTF1} Erd\'{e}lyi A,  Magnus W,  Oberhettinger F, and Tricomi F G  1953
{\em Higher Transcendental Functions} (New York: McGraw--Hill), Vol. 1. 
\bibitem{Usha} Usha Devi A R, Prabhu R, and Uma M S 2006
Non-classicality of photon added coherent and thermal radiations
{\em Eur. Phys. J. D}  {\bf 40} 133 
\bibitem{Asboth} Asb\'oth J K,  Calsamiglia J, and  Ritsch H 2005
Computable Measure of Nonclassicality for Light {\em Phys. Rev. Lett.} {\bf  94} 173602
\bibitem{P91} Marian P 1991 Higher-order squeezing properties and correlation functions 
for squeezed number states {\em Phys. Rev A} {\bf 44}  3325  
\bibitem{Shimony} Shimony A 1995 Degree of entanglement {\em Ann. N. Y. Acad. Sci.}  {\bf 755} 675 
\bibitem{TZ} Takagi R and  Zhuang Q 2018 Convex resource theory of non-Gaussianity 
{\em Phys. Rev A} {\bf 97} 062337 
\bibitem{Tav} Safaeian O and Tavassoly M K 2011 Deformed photon-added
nonlinear coherent states and their non-classical properties, {\em J.
Phys. A: Math. Theor. } {\bf 44} 225301
\bibitem{Moja} Mojaveri B,  Dehghani A, and Mahmoodi S, 2014  New class of generalized photon-added 
coherent states and some of their nonclassical properties {\em Phys. Scr.} {\bf  89}  085202 
\bibitem{Hussin} Dey S and  Hussin V 2016 Noncommutative q-photon-added coherent states 
{\em Phys. Rev A} {\bf 93}, 053824


\end{thebibliography}
\end{document}